\def\roughly#1{\mathrel{\raise.3ex\hbox{$#1$\kern-.75em
\lower1ex\hbox{$\sim$}}}}

\def\IR{\relax{\rm I\kern-.18em R}}
\def\IQ{\relax{\rm I\kern-.18em Q}}
\def\IF{\relax{\rm I\kern-.18em F}}
\def\IH{\relax{\rm I\kern-.18em H}}
\font\cmss=cmss10 \font\cmsss=cmss10 at 7pt
\def\IZ{\relax\ifmmode\mathchoice
{\hbox{\cmss Z\kern-.4em Z}}{\hbox{\cmss Z\kern-.4em Z}}
{\lower.9pt\hbox{\cmsss Z\kern-.4em Z}}
{\lower1.2pt\hbox{\cmsss Z\kern-.4em Z}}\else{\cmss Z\kern-.4em Z}\fi}

\def\sltwoz{SL(2,\IZ)}

\documentstyle[prl,aps]{revtex}
\begin{document}
\twocolumn[\hsize\textwidth\columnwidth\hsize\csname@twocolumnfalse%
\endcsname

\rightline{McGill-98/36}
\rightline{{\tt cond-mat/9812396}}
\draft

\title{On the Implications of Discrete Symmetries for the $\beta$-function\\
of Quantum Hall Systems}

\author{C.P.~Burgess${}^a$ and C.A.~L\"utken${}^b$}

\address{${}^a$ Physics Department, McGill University,
3600 University Street, Montr\'eal, Qu\'ebec, Canada H3A 2T8.
\\
${}^b$ Department of Physics, University of Oslo, P.O. Box 1048 Blindern,
N-0316 Oslo, Norway.}

\maketitle

\begin{abstract}
{
We argue that the large discrete-symmetry group of quantum Hall systems is insufficient
in itself to determine the complete $\beta$-function for the scaling of the conductivities,
$\sigma_{xx}$ and $\sigma_{xy}$. We illustrate this point by showing
that a recent ansatz for this function is one of a many-parameter family. A clean
prediction for the delocalization exponents for these systems 
therefore requires the specification
of more information, such as past proposals that the $\beta$-function is either
holomorphic or quasi-holomorphic in the variable 
$z = (\hbar/e^2)(\sigma_{xy} + i \sigma_{xx})$.
}
\end{abstract}
\pacs{PACS numbers:  }
]

%
%

\section{Introduction}

It has been conjectured \cite{BL} that the large group of discrete symmetries 
enjoyed by quantum Hall systems \cite{LR,KLZ} might permit the complete
determination of the $\beta$-function of these materials. This $\beta$-function
describes the renormalization-group (RG) flow of the conductivities $\sigma_{xx}$
and $\sigma_{xy}$ in the delocalization-scaling theory of these systems. 
The question addressed in ref.~\cite{BL} was whether a few physically well-motivated
constraints on the $\beta$-function, exploiting available data both from the 
weakly and the strongly coupled domains of the system, suffice to completely 
fix the full non-perturbative form of the $\beta$-function in an 
appropriate scheme, without actually deriving it from a microscopic theory. 

More precisely, the main idea was that the consistency of the RG flow with 
the discrete symmetry group --- which relates large and small 
$\sigma_{xx}$ --- might be sufficient to completely determine the 
$\beta$-function if combined with the known weak-coupling 
results in the asymptotic domain of large $\sigma_{xx}$. Such a
construction would specify, among other quantities, 
the universality class (and critical exponents) of the quantum critical
``anyon delocalization'' points which are believed to exist in this system.
Since this goal was largely thwarted by the existence of more than one
$\beta$-function which satisfied these conditions,
the main focus of ref.~\cite{BL} is the establishment of further conditions 
which would uniquely determine a solution. 

There has been a resurgence of interest in this topic of late, with the recent
appearance of two ans\"atze \cite{DOLAN,TAN} for the nonperturbative form of 
this exact $\beta$-function. In this letter we add our own multiple-parameter
ansatz to this list. We revive this ansatz not because we believe it to be the final
word on this subject, but rather to emphasize that the determination of the 
$\beta$-function requires more than simply symmetry and asymptotic information,
as well as to clarify the relation of these ans\"atze with the conditions outlined 
in ref.~\cite{BL}. 

In ref.~\cite{BL} it was the necessity for extra information which drove the formulation 
of quasi-holomorphy, while in ref.~\cite{DOLAN} it is holomorphy
which is invoked for the same reasons. It would be interesting to understand 
what this information might be for the ansatz of ref.~\cite{TAN}, which 
is sufficiently narrow to produce specific values for the delocalization 
critical exponents. It is otherwise difficult to know why this ansatz 
should be preferable over others which may also do so.

We present our ideas in the following way. First, we briefly recap the
symmetry and asymptotic conditions which we demand of the 
$\beta$-function. In so doing we expand on the consequences of 
holomorphy, in order to make contact with ref.~\cite{DOLAN}.
Next, we describe our ansatz, which contains that of ref.~\cite{TAN}
as a special case. We use this ansatz to explore in more detail the
form taken by $\beta$ at large $\sigma_{xx}$ and near the critical
points. 

\section{Symmetries and Asymptotics}

The precise form of the constraints identified in ref.~\cite{BL} are:

\vskip 0.3cm
{\it 1. Asymptotics:}    
The $\beta$-function which is predicted by the effective sigma-model of 
weak localization in a magnetic field (see ref.~\cite{EFETOV} for a comprehensive review) 
may be explicitly computed for large $\sigma_{xx}$, since this corresponds
to weak coupling in the sigma-model. Writing $\sigma_{xy} + i\sigma_{xx} 
= (e^2/\hbar)\; z$ where $z = x + iy$ and $\bar z = x - i y$, we have:
\begin{equation}
\beta^z  = \frac{dz}{dt} = 
b_0 + \frac{b_1}{y} + \frac{b_2}{y^2} + \dots + (q, \bar q-{\rm expansion}) .
\label{betapert}
\end{equation}
Here $t$ is a logarithmic scale parameter, $q = \exp(2 \pi i z)$,
$\bar q = \exp(-2\pi i\bar z)$, and the ``$q,\bar q$-expansion''
refers to the leading non-perturbative (dilute instanton gas) corrections 
to the perturbative loop expansion in $1/y$.  

The coefficients $b_i$ of the perturbative sigma-model $\beta$-function are 
known through six-loop order ($i \le 4$) \cite{HIKAMI}, with values:
\begin{equation}
b_{0} = 0,\quad b_1 = - \; \frac{i}{2 \pi^2}, \quad b_2 = 0,
\quad b_3 = - \; \frac{3i}{8 \pi^4}, \quad b_4 = 0 .
\label{betacoeffs}
\end{equation}
The even powers of $1/y$ vanish in this renormalization scheme
(to the order known). Note, however, that only the leading
non-vanishing coefficient $b_1$ is scheme independent.
By contrast, only the leading term in the instanton expansion 
is available \cite{PRUISKEN}, and is proportional to the anti-instanton 
result: $y^k \bar q$, for $k$ positive.

There are three senses in which the $\beta$-function might be asked to
agree with eqs.~(\ref{betapert}) and (\ref{betacoeffs}). 

\begin{enumerate}
\item {\it Weak Agreement:}
The weakest condition simply requires agreement with the scheme-independent
perturbative part of these results --- {\it i.e.} with only $b_0$ and $b_1$. 
\item {\it Strong Agreement:}
A stronger requirement is agreement with the entire perturbative expansion,
as computed in the sigma-model. 
\item {\it Very Strong Agreement:}
The strongest requirement would be agreement with all known terms, {\it including}
the dilute instanton gas expansion.  None of the ans\"atze which have been 
proposed to date satisfy this condition, since they do not properly reproduce 
the power of $y$ in the leading instanton result.  We shall not here focus much
attention on this condition, however, since existing instanton calculations for 
disordered systems are performed using the replica trick, which is known to 
sometimes fail even for systems where the perturbative results they give are 
accurate \cite{REPLICAFAIL}.  
\end{enumerate}

\vskip 0.3cm
{\it 2.  Scaling:}
We require the $\beta$-function to share the (real) analytic structure
to be expected of any RG flow. That is, since $\beta^z$ describes
the differential elimination of degrees of freedom at a particular
scale, it does not contain singularities, apart from those
places where the number of relevant degrees of freedom change.
We therefore demand that the $\beta$-function should be non-singular 
throughout the upper half of the complex $z$-plane ($\IH: {\rm Im} z > 0$). 
The continuous (second order) quantum critical delocalization transitions 
in this system (see ref.~\cite{SGCS} for a review) are then identified 
with the zeros of $\beta$.  From eqs.~(\ref{betapert}) 
and (\ref{betacoeffs}) it is clear that $\beta$ also goes
to zero as $z \to i\infty$.

Because $\sigma_{xy} = (e^2/\hbar)\; x$ enters the sigma model as the coefficient
of a topological term, $\beta$ cannot depend on $x$ to any order in perturbation
theory, implying that the leading asymptotic behaviour of $\beta$ is a function
of $y$ alone.  Thus, unless all $b_i$ except $b_0$ vanish (as happens in 
systems with unbroken complex supersymmetry), agreement with the
sigma-model large-$y$ form precludes the $\beta$-function 
being a complex analytic (holomorphic) function of $z$ only.

The exploitation of scaling ideas and data are at the very heart of the 
``phenomenological'' approach proposed in ref.~\cite{LR}, and further
pursued in ref.~\cite{BL}.  It is universality which allows us to entertain the idea 
that the $\beta$-function could be determined up to asymptotics
by macroscopic scaling properties.  Conversely, should the $\beta$-function 
be determined, it need not shed much light on the nature of the microscopic physics 
responsible for the scaling laws in this system.
It can, however, show that all the scaling data, as well as the 
phase diagram and Hall quantization, are encoded in the low-energy 
effective theory as a global discrete symmetry of the (complexified)
Kramers-Wannier type.  It is this alleged symmetry which provides 
the final physically-motivated constraint on the $\beta$-function. It
also is what endows our conjecture with most of its power, relating 
properties which are perturbative within the sigma-model context
to those which are not.

\vskip 0.3cm
{\it 3.  Automorphy:}    
As explained at length elsewhere \cite{LR}, the observed 
``superuniversality'' of the critical exponents in the hierarchy 
of delocalization transitions that take place in the quantum Hall system
was the original motivation for conjecturing that the low-energy 
theory respects, in the fully spin-polarized case, a global discrete 
symmetry $\Gamma = \Gamma_0(2)$. ($\Gamma_0(2)$, which
is also denoted $\Gamma_{\rm T}(2)$ in refs.~\cite{BL,LR}, is a well-known
sub-group of the modular group, $\sltwoz$. )
The mathematical fact that this subgroup automatically ensures 
Hall quantization on odd-denominator fractions (when $\sigma_{xx}\rightarrow 0$) 
is also encouraging. Independent arguments arose at about the same time
in the form of  the ``law of corresponding states'',
from a mean-field treatment of the microscopic theory \cite{KLZ}. 

The modular group (and its subgroups) act on the complex conductivity
as special M\"obius transformations: $\gamma(z) = (az + b)/(cz + d)$ 
where $a,b,c,d$ are integers satisfying $ad - bc = 1$. The subgroup
$\Gamma_0(2)$ is defined by the additional condition that $c$ be
even. 

If such a symmetry is present at low energy the $\beta$-function must respect 
it in a very specific sense. A function $f(z,\bar z)$ is called automorphic of 
weight $(u,v)$ under $\Gamma$  {\it iff} it transforms like a generalized tensor:
\begin{eqnarray}
f(\gamma(z),\gamma(\bar z)) &=& \left(\frac{d\gamma}{dz}\right)^{-u/2}
\left(\frac{d\bar\gamma}{d\bar z}\right)^{-v/2} f(z,\bar z) \nonumber\\
&=& (cz+d)^u (c\bar z+d)^v f(z,\bar z)
\end{eqnarray}
for every $\gamma$ in $\Gamma$.

It was shown in ref.~\cite{BL} that if the RG commutes with $\Gamma$, then 
the physical (contravariant) $\beta$-function $\beta^z$ is a {\it negative} 
weight $(-2,0)$ function, while the complex conjugate function, $\beta^{\bar z}$, 
has weight $(0,-2)$. Similarly, given a metric $G_{ij}$, the covariant 
$\beta$-function, $\beta_i = G_{ij} \beta^j$ must have positive 
weights: $\beta_z \sim (2,0)$  and $\beta_{\bar z} \sim (0,2)$.

\vskip 0.5cm
Because the constraints {\it 1} through {\it 3} are extracted from experimental 
data and/or general knowledge about scaling and perturbation theory, 
they would seem to be a reasonable starting point for the search for
the exact quantum-Hall $\beta$-function. Since we display many
solutions to these conditions below they cannot be sufficient in themselves
to uniquely determine the result. 

Before presenting these solutions we pause to discuss the holomorphy
and quasiholomorphy assumptions.

\section{Holomorphy}

A natural guess for $\beta^z$ is that it is a holomorphic (or anti-holomorphic) 
function: $\beta^z = \beta^z(z)$ (or $\beta^z = \beta^z(\bar z)$), a proposal
recently revived in ref.~\cite{DOLAN}. This is a
very predictive ansatz because it permits the use of powerful results from
complex analysis. As discussed above, this ansatz is inconsistent with 
even the weak form of the sigma-model behaviour at large $y$, and so
it necessarily implies the breakdown of this sigma-model description of
weak localization in magnetic fields. The purpose of the present section is
to establish that a holomorphic (or anti-holomorphic) $\beta$-function {\it must}
also have a singularity somewhere in $\overline\IH = \IH\cup\IQ\cup\infty$, where $\IQ$
are the rational numbers. One could conceivably tolerate a pole on the
real axis, but it is then difficult to obtain an acceptable flow.  
For these two reasons the holomorphic option was rejected in ref.~\cite{BL}.

A particularly useful fact for any meromorphic
function $f$ of weight $(k,0)$ with respect to $\Gamma_0(2)$, 
relates the `index' of its zeros and poles within a fundamental domain 
$\overline\IF$ of $\overline\IH$ \cite{FOOTNOTE1}:
\begin{equation}
n_\infty + n_0 + \frac{n_*}{2} + \sum_p n_p = \frac{k}{4} .
\label{index}
\end{equation}
Here $n_p$ is the leading power of $z-z_p$ which appears in a Laurent
expansion of $f$ about the pole or zero at $z_p$ in the interior or boundary of
$\overline\IF$. $n_*$ is the same quantity in the expansion of $f$ about the
fixed point, $z_* = (1 + i)/2$, of the group $\Gamma_0(2)$. Similarly,
$n_\infty$ is the leading power in a Laurent 
expansion of $f$ in powers of $q$ about
$z = i \infty$, while $n_0$ counts the leading power of $\tilde q = \exp(-i \pi/z)$
in an expansion of $z^k f(z)$ in powers of $\tilde q$ 
about $z = 0$ \cite{RANKIN,KOBLITZ}. 

Eq.~(\ref{index}) implies, in particular, that no weight $(-2,0)$ function like
$\beta^z$ can be holomorphic without acquiring singularities somewhere in 
$\overline\IH$. 
For example, as was observed in ref.~\cite{RITZ}, the $\beta$-function
for the Seiberg-Witten $N=2$ supersymmetric $SU(2)$ gauge theory is the
unique weight $(-2,0)$ function (up to normalization) which has a simple
zero only at $z_*$ (and its images under $\Gamma_0(2)$) and which 
approaches a constant as $z \to i \infty$. The complex quantity $z$ in this
case is related to the gauge coupling ($g$) and vacuum angle ($\theta$) by:
$z = (\theta /2 \pi) + (4 \pi i/g^2)$. 
Eq.~(\ref{index}) forces $\beta(z)$ to have a simple pole at $z = 0$ 
(as well as the other integers on the real axis), 
corresponding to the infinite-coupling limit in the sigma-model.

Similar considerations apply if the quantum Hall $\beta$-function 
were to be holomorphic, and in particular it would also
be singular somewhere. A simple proposal is to choose $\beta^z$
or $\beta^{\bar z}$ to be proportional to the Seiberg-Witten
$N=2$ supersymmetric $\beta$-function. Unfortunately, the flow
in this case is {\it repelled} by the odd-denominator
fractions on the real axis, instead flowing towards $z_*$,
which is an attractive fixed point, with no irrelevant directions.
Clearly this flow cannot describe the second-order
transitions of the quantum Hall systems.

One might imagine making more complicated choices, such as to
force the $\beta$-function to have simple
zeroes ({\it i.e.} $n_\infty = n_* = 1$) both at $i\infty$ and 
$z_*$, with no poles or singularities elsewhere for nonzero
$\sigma_{xx}$ --- thus making it at least qualitatively similar to the 
perturbative sigma-model result. Such a condition must
have a double pole at $z=0$. We now argue that
{\it any} holomorphic $\beta$-function having a simple
zero at $z=z_*$ cannot have both a relevant and irrelevant
direction there, giving an unacceptable flow. 

To establish this result proceed as follows. The critical exponents
are related to the derivative of the $\beta$-function at its zeroes,
and a simple argument shows that holomorphy dictates that these 
must have the same sign. This is because the the matrix of 
derivatives for holomorphic $\beta^z$ necessarily has the
following form:
\begin{equation}
\label{samesign}
\pmatrix{ {d\beta^z \over dz} & {d\beta^z \over d \bar z} \cr
{d\beta^{\bar z} \over dz} & {d\beta^{\bar z} \over d \bar z} \cr}
= \pmatrix{ B & 0 \cr 0 & \bar B \cr},
\end{equation}
from which we see that the product of the eigenvalues of this
matrix is $B\bar B \ge 0$, implying: ({\it i}) both eigenvalues have the
same sign; or ({\it ii}) one (or both) is zero. Neither of these cases
describes the observed flow near the quantum Hall critical points.

As observed in refs.~\cite{DOLAN,LATLUT}, 
qualitatively acceptable flows are 
obtained by moving the pole of the $\beta$-function 
to the fixed point $z_*$, in which case the $\beta$-function 
can approach a constant as $z\rightarrow i\infty$ and $z\rightarrow 0$. 
The simplest such $\beta$-function --- which has $n_* = -1$
and all others zero --- turns out to be just the 
inverse of the holomorphic weight $(2,0)$ 
function ${\cal E}(z)$ defined below.
As discussed in ref.~\cite{DOLAN}, the pole makes it problematic 
to identify the universal critical exponents at $z_*$, and 
ref.~\cite{DOLAN} instead explicitly exhibits the flow near this point 
in order to make comparisons with the data.

\section{Quasi-holomorphy}

The alternative followed in ref.~\cite{BL} was to 
start with the observation that eq.~(\ref{index})
is much more kind in its implications for the covariant function, $\beta_z$,
than it is for $\beta^z$. This is because $\beta_z$ transforms under
$\Gamma_0(2)$ as an automorphic function of {\it positive} weight $(k = 2)$. 
In fact, for $\Gamma_0(2)$ --- but not for $\sltwoz$ --- there is a 
unique (up to normalization) holomorphic $(2,0)$ function ${\cal E}(z)$, 
which is nowhere singular on $\overline\IH$.  It can be expressed in terms of 
the famous modular discriminant function $\Delta = q \prod(1 - q^n)^{24}$
which generates the holomorphic (but {\it not} automorphic) Eisenstein 
function:  
\begin{equation}
E_2(z) = \frac{1}{2\pi i} \partial_z \log \Delta 
=  1 - 24 \sum_{n=1}^{\infty} \frac{nq^n}{1 - q^n},
\end{equation}
as follows: 
\begin{equation}
{\cal E}(z) =  2 \, E_2(2z) - E_2(z) = 
1 + 24 \; \sum_{n=1}^{\infty} \frac{nq^n}{1 + q^n}.
\label{holo}
\end{equation}
${\cal E}(z)$ transforms as a weight $(2,0)$ function with respect
to $\Gamma_0(2)$ even though $E_2(z)$ does not.
Unfortunately, since ${\cal E}(z)$ does not vanish as $z \to i \infty$
it is inconsistent with the perturbative expression, eq.~(\ref{betapert}).

The inconsistency between ${\cal E}(z)$ and the perturbative result
in powers of $1/y$ motivates the search for more general 
weight $(2,0)$ quantities which are not holomorphic but have the more
general form: $1/y + g(z)$, for holomorphic $g(z)$. Such functions
were called quasi-holomorphic in ref.~\cite{BL}. 

The most general quasi-holomorphic $(2,0)$ form which is nowhere singular
in $\IH$ is a linear combination of ${\cal E}(z)$ and:
\begin{eqnarray}
{\cal H}(z,\bar z) &=& \frac{1}{\pi \, y} + \frac{2}{3} \, \Bigl[ E_2(2z)
- E_2(z) \Bigr] \nonumber\\
&=& \frac{1}{\pi \, y} + 16 \; \sum_{n=1}^{\infty} \frac{nq^n}{1 - q^{2n}}.
\label{hecke}
\end{eqnarray}
Both ${\cal E}$ and ${\cal H}$ vanish at $z_* = (1+i)/2$.
It is the unlikely existence of this quasi-holomorphic ``Hecke'' function
\cite{HECKE} which makes the idea of quasi-holomorphy useful.
Further motivations for restricting attention to quasi-holomorphic building 
blocks are discussed in ref.~\cite{BL}.

\vskip 0.3cm
{\it 4. Quasi-holomorphy:} 
Ref.~\cite{BL} therefore proposed that $\beta_z$ is quasi-holomorphic. 
The unique quasi-holomorphic choice which is consistent with the weak
form of agreement with sigma-model perturbation theory then is (if
$G_{ij} \to 1$ as $z \to i \infty$):
\begin{equation}
\beta_z = \frac{i }{2 \pi} \; {\cal H}  \; .
\label{qholo}
\end{equation}
With this proposal, the scheme dependence of $\beta^z$ enters through
the definition of the metric, $G_{ij}$. 

\section{More Ans\"atze}

If the $\beta$-function is not holomorphic then the 
previously-mentioned conditions are insufficient to completely
pin it down. To establish this point, we now exhibit many 
more ans\"atze which satisfy all but the very-strong 
asymptotic condition. 

Since the ratio of any two $(0,-2)$ functions is a $(0,0)$ function,
any $(0,-2)$ function $\beta^{\bar z}$ can be written as: 
\begin{equation}
\beta^{\bar z} = \frac{i }{2 \pi} \;W \; R,
\label{genform}
\end{equation}
where, $R(z,\bar z)$ is a weightless (weight $(0,0)$) function to be
specified and, in the spirit of quasi-holomorphy, we choose to write the 
weight $(0,-2)$ factor as  $W = {\cal H}/{\cal D}$ with:
\begin{equation}
{\cal D} = \vert {\cal E}\vert^2 + a\pi^2 \vert {\cal H}\vert^2 +  
\pi \, b\, {\cal E }\overline{\cal H} 
+ \pi  \, c\, {\cal H} \overline{\cal  E} + \frac{d}{y^2} .
\label{Ddef}
\end{equation}
Here $a,b,c$ and $d$ are constants, and $\cal D$ transforms under $\Gamma_0(2)$ 
as a weight $(2,2)$ function.

In order to make the invariance
of $R$ with respect to $\Gamma_0(2)$ explicit, it is convenient to
change variables from $z$ to $f = - \vartheta_3^4(z) \, \vartheta_4^4(z)/
\vartheta_2^8(z)$, and write $R = R(f,\bar f)$. This may always be
done since this $f $ plays the same role
for $\Gamma_0(2)$ as Klein's famous $j$-function does for the full
modular group.  In particular, it is invariant under $\Gamma_0(2)$
and uniquely labels every point in the fundamental domain $\overline\IF$
of the group ({\it i.e.} it is a one-to-one map of $\overline\IF$
onto the complex sphere).

So far we have made no assumptions beyond automorphy. 
In choosing our ansatz for $a,b,c,d$ and $R$ our guidance is 
(strong) agreement with the large-$y$ limit, as well as 
requiring a zero of $\beta^{\bar z}$ at $z_*$, and
the absence of singularities and zeros elsewhere in $\IH$. 

\vskip 0.3cm
{\it Ansatz 1:} 
The simplest case is to choose $d \neq 0$, in which case ${\cal D} > 0$ throughout $\IH$ 
so long as $b$ and $c$ are sufficiently small.  In this case all assumptions are 
satisfied with the choice $R = 1$. Thus:
\begin{eqnarray}
\beta^{\bar z}(z,\bar z) &=& \frac{i}{2\pi} \;\frac{ {\cal H} (z,\bar z)}
{{\cal D}(z,\bar z)}\nonumber\\
&=& \frac{i}{2\pi^2\, y} \; \left( 1 + \frac{b + c}{y} + \frac{a+d}{y^2}\right)^{-1}
+ O(q,\bar q).
\label{ansatzone}
\end{eqnarray}

Notice that $b_2 = 0$ in agreement with eq.~(\ref{betacoeffs})
if we set $b = - c$ in the ansatz, in which case $b_{2n}=0$ for all $n$.  
Similarly $b_3$ is properly reproduced if $a+d = - 3/(4 \pi^2) \approx - 0.076$. 
This ansatz then predicts all other terms in the perturbative series:
\begin{equation}
\beta^{\bar z}_{\rm pert} = \frac{i}{2\pi^2\, y} \; 
\left( 1 - \frac{3}{4 \pi^2 y^2} \right)^{-1} 
\label{pertpredn}
\end{equation}
which gives the following coefficients $b_i$:
\begin{equation}
b_{2n} = 0, 
\qquad b_{2n+1} = - \; \frac{3^n i}{2^{2n+1}\pi^{2n+2}}.
\label{fullbetacoeffs}
\end{equation}

In principle, the values of $a,b,c$ and $d$ can be separately 
extracted by comparison with the leading non-perturbative terms 
proportional to $q$ and $\bar q$. Writing $D =  1 - 3/(4 \pi^2 y^2) $
this gives:
\begin{eqnarray}
\beta^{\bar z}_{q,\bar q} &=& {8 i \over \pi D} \; \left[1 -
{1 \over 2 \pi y D} \; \left( 3 + 2\pi \, c + { 3 b + 2 \pi \, a \over y}
\right) \right] \; q \nonumber\\
&& \qquad  - \; {4 i \over \pi^2 \, y D^2} \; \left(3 + 2 \pi \, b 
+ {3 c + 2 \pi \, a \over y} \right) \; \bar q .
\end{eqnarray}
Notice, however, that the leading powers are $y^0 q$ and 
$y^{-1} \bar q$, which does not agree with ref.~\cite{PRUISKEN},
(who finds a positive power of $y$ premultiplying $q$).

The ansatz of eq.~(\ref{ansatzone}) has a simple zero 
at the fixed point, $z_* = (1 + i)/2$.
The critical exponents at this point are found by diagonalizing the 
matrix of derivatives of the $\beta$-function at this point. 
Using ${\cal E}(z) \approx - 6.10 i (z - z_*) + O((z - z_*)^2)$
and ${\cal H}(x,y) \approx - 3.69i (x - x_*) + 2.41 (y - y_*) + \dots$,
we find  the localization length exponent $\nu \approx d / 0.147$ and 
irrelevant exponent (see ref.\cite{BODO1} 
for definitions and a review of experimental results) $y 
\approx - 0.096/d$ \cite{FOOTNOTE2}. Choosing the parameter
$d \approx 0.34$ puts the prediction for $\nu$ in agreement
with experimental results $\nu_{\rm exp} = 2.3 \pm 0.1$ \cite{KOCH}   
($2.4 \pm 0.2$ \cite{WEI}). 

This choice for $d$ permits an absolute prediction from this
ansatz for the irrelevant exponent: $y \approx -0.29$, which does not seem 
to reproduce the results of numerical simulations, which 
give $\nu_{\rm num} = 2.35 \pm 0.03$ \cite{BODO2} , 
and $y_{\rm num} = - 0.38 \pm 0.02$ \cite{CHALKER} 
($- 0.42 \pm 0.04$ \cite{BODO3}). 

\vskip 0.3cm
{\it Ansatz 2:} 
More complicated ans\"atze are also possible. For example,
if $d = 0$, then $R$ must be chosen to vanish at $z_*$ 
in order to cancel the pole in ${\cal H}/{\cal D} \vert_{d=0}$. This is easily
arranged since $f- 1/4$ has its only (double) zero at this point.

This type of ansatz contains the one proposed in ref.~\cite{TAN}
as the special case  $b = -c = A$ and $a = -A^2$, with $R$ given
by the rational function $R = \left(Q - 1/2 \right)/Q$, and 
$Q = f + \bar f + 2(f - \bar f)/(\pi A)$. The value of the parameter $A
\approx 0.623$ is chosen to ensure the cancellation of the pole at $z_*$.

Although for holomorphic functions such a rational form for
$R$ follows on general grounds \cite{RANKIN,KOBLITZ},
we are not aware of any similar result for the nonholomorphic functions
considered here.

This particular ansatz also does not agree, in the strong sense, with the 
sigma-model result at large $y$. This is because $R = 1 + O(q,\bar q)$
as $z \to i \infty$, and so its prediction for the perturbative 
$\beta$-function is the same as the perturbative part of
eq.~(\ref{ansatzone}), with $b+c=0$ and
$a+d=a=-A^2 \approx -0.388$. This clearly {\it disagrees} with the
perturbative theory already at order $O(1/y^3)$. If only agreement in 
the weak sense is desired, then there is no reason to set $b+c = 0$, 
exhibiting this ansatz as one of a several-parameter family. 

Putting aside agreement with sigma-model perturbation theory,
the predictions for the critical exponents obtained from this ansatz 
in ref. \cite{TAN} become $\nu = 2.12$ and $y = - 0.31$, which
is consistent (within the roughly 10\% errors) with experimental 
scaling data, but has difficulty with the numerical simulations of the quantum 
Hall system \cite{BODO2,CHALKER}. 

As was already pointed out in ref.\cite{TAN}, this
ansatz varies as $y^0 q$ and so, like all of the previous
ans\"atze, disagrees with the leading nonperturbative terms 
predicted by the sigma-model. One might wonder if the
requirement of agreement with the sigma model in the very 
strong sense could itself be the remaining condition which
uniquely determines the form of $\beta^z$. 
Leading instanton correction of the form $y^k q$ for $k=1$ or $k=2$,
as required by the very strong asymptotic condition if the sigma-model 
instanton calculation \cite{PRUISKEN} is taken seriously, can be
achieved within the framework of the ans\"atze considered here by 
adding terms like $F = y^2{\cal E}/f\rightarrow y^2q + \dots$ or 
$G = y^2{\cal H}/f \rightarrow yq + \dots$ to the product $WR$.
Both $F$ and $G$ are weight $(0,-2)$ functions with simple zeros at 
$z_*$ (and at $z=0$), and so would change the values which
are inferred for the critical exponents.

It would be encouraging to hope that concentration on this
condition, or a solution to the consistency conditions 
for the metric $G_{ij}$ discussed in ref.\cite{BL}, 
might lead to further progress in finding
sufficient conditions for the determination of the exact
scaling properties of the quantum Hall system.

\section{Acknowledgements}
\vspace{0.1cm}
This research has been supported in part by NSERC (Canada), FCAR (Qu\'ebec)
and the Norwegian Research Council.  We thank Dan Arovas, Brian
Dolan and Cyril Furtlehner for useful discussions.



\begin{references}
\bibitem{FOOTNOTE1}
A fundamental domain $\IF$ is the quotient of $\IH$ by the 
symmetry group $\Gamma$, $\IF = \IH/\Gamma$, while the 
compactified fundamental domain, $\overline\IF$ includes 
the ``cusp points'' on the real line ($z=0$ in our case where 
$\Gamma = \Gamma_0(2)$).
\bibitem{FOOTNOTE2}
We here follow standard usage, denoting by $y$ the irrelevant index
$1/\nu_{\rm irrel}$. This should not be confused with $y = \Im z$ as
used elsewhere in the text.
\end{references}
\end{document}